\documentstyle[a4,12pt,amssymb,times]{article}
\newcommand{\be}{\begin{equation}}
\newcommand{\ee}{\end{equation}}
\newcommand{\bea}{\begin{eqnarray}}
\newcommand{\eea}{\end{eqnarray}}

\def\pr{\pi_{\rho}}
\def\pf{\pi_{\phi}}
\def\pz{\pi_0}
\def\p1{\pi_1}
\def\b{\beta}
\def\l{\lambda}
\def\f{\phi}
\def\r{\rho}
\def\e{\eta}
\def\t{\theta}
\def\pmas{\partial_+}
\def\pmen{\partial_-}
\def\v{\vskip}

\begin{document}
\begin{titlepage}
\begin{flushright}
{\tt FTUV/98-53\\
     IFIC/98-54\\
     hep-th/9807003}
\end{flushright}

\vspace*{0.5cm}

\begin{center}
{\Large\bf B\"acklund transformations in 2D dilaton gravity\footnote{
Work partially supported by the 
{\it Comisi\'on Interministerial de Ciencia y Tecnolog\'{\i}a}\/ 
and {\it DGICYT}.}}
\\[0.5cm]
D. J. Navarro\footnote[2]{E-mail address: dnavarro@lie.uv.es} and 
J. Navarro-Salas\footnote[3]{E-mail address: jnavarro@lie.uv.es}.
\\[0.5cm]
{\footnotesize
       Departamento de F\'{\i}sica Te\'orica and
       IFIC, Centro Mixto Universidad de Valencia-CSIC.\\
       Facultad de F\'{\i}sica, Universidad de Valencia,
       Burjassot-46100, Valencia, Spain.}
\end{center}

\bigskip

\begin{abstract}
We give a B\"acklund transformation connecting a generic 2D dilaton gravity
theory to a generally covariant free field theory. This transformation
provides an explicit canonical transformation relating both theories.
\end{abstract}

\v8cm

\begin{flushleft}
PACS number(s): 04.60.Kz, 04.60.Ds
\end{flushleft}

\end{titlepage}
\newpage

Motivated by the two-dimensional model of black holes dynamics introduced
by Callan, Giddings, Harvey and Strominger (CGHS) \cite{cghs}, a lot of works
in 2D dilaton gravity has been developed from different viewpoints. A crucial
property to understand the CGHS model is given by the fact that it can be
mapped, via an off-shell canonical transformation, into a theory of free fields
with a Minkowskian target space \cite{cangemi}.
This, in turns, implies that theory can be
quantized using different approaches to deal with the anomalies \cite{cangemi,
cavaglia}. It has been proved in Ref. \cite{cruz1} (see \cite{cruz2} for
details) that this property of the CGHS model is also valid for a generic model
of 2D dilaton gravity. Based on properties of the classical equations of motion
it was shown that there exist a canonical transformation converting a generic
model into a free field theory with a Minkowskian target space. However the
explicit form of the transformation is unknown except for those models which
can be explicitly solved \cite{filippov}.\\

The aim of this letter is to provide a more explicit form for the canonical
transformations and bypass the problem of solving the classical equations of
motion. To this end we shall introduce a different perspective to that used in
\cite{cruz1,cruz2}. The idea is to construct a B\"acklund transformation 
relating the equations of motion of a 2D dilaton gravity model to free field
equations. To obtain the B\"acklund transformation we shall
also demand that the Hamiltonian and momentum constraints of the
dilaton-gravity theory are mapped into those of a generally covariant free
field theory. With this requirement the B\"acklund transformation can be
promoted to a canonical transformation.\\

Our starting point is the action functional describing a 2D dilaton gravity
model
\be
\label{action2D}
S=\int d^2x \sqrt{-g} \left( R\phi + 4\l^2 V(\f) - \frac{1}{2} (\nabla f)^2
\right) \, ,
\ee
where $V(\f)$ is an arbitrary function of the dilaton field and $f$ is a scalar
matter field. The above expression represent a generic model because one can
get rid of the kinetic term of the dilaton by a conformal reparametrization of
the fields and bring the action into the form (\ref{action2D}). In conformal
gauge $ds^2=-e^{2\r}dx^+dx^-$, the equations of motion derived from the action
(\ref{action2D}) are
\bea
\label{eq1}
2\pmas\pmen\r + \l^2 \frac{d}{d\f}V(\f) e^{2\r} & = & 0 \, , \\
\label{eq2}
\pmas \pmen \f + \l^2 V(\f) e^{2\r} & = & 0 \, , \\
\label{eq3}
\pmas \pmen f & = & 0 \, , \\
\label{eq4}
-\partial_{\pm}^2 \f + 2\partial_{\pm} \f \partial_{\pm} \r -\frac{1}{2}
(\partial_{\pm}f)^2 & = & 0 \, .
\eea
By a rather involved manipulation of these equations it was shown in 
\cite{cruz1,cruz2} that, irrespective of the form of the potential, the
solutions define a canonical transformation mapping the theory (\ref{action2D})
into a free field theory with constraints $C_{\pm} = \pm \frac{1}{2} (H \pm P)$
taking the form
\be
C_{\pm} = \Pi_{\pm} X^{\pm \prime} \pm \frac{1}{4} (\pi_f \pm f^{\prime})^2
\, ,
\ee
where $(\Pi_{\pm},X^{\pm})$ and $(\pi_f,f)$ are cano\-ni\-cally conjugate
variables and $H$ and $P$ are the Hamiltonian and momentum constraints.
Obviously the pure gravity and matter sectors are separately equivalent to
free fields. From now we shall restrict our analysis to the pure
dilaton-gravity sector. A further linear canonical transformation \cite{kuchar}
\bea
2\Pi_{\pm} & = & -(\pi_0 + \pi_1) \mp (r^{0\prime} - r^{1\prime}) \, , \\
2X^{\pm\prime} & = & \mp (\pi_0 - \pi_1) - (r^{0\prime} + r^{1\prime}) \, ,
\eea
converts finally the constraints into those of a free field theory with a
Minkowskian target space
\be
\label{cfree}
C_{\pm} = \pm \frac{1}{4} \left[ (\pz \pm r^{0\prime})^2 - 
(\p1 \mp r^{1\prime})^2 \right] \, .
\ee

As we have already mentioned it is in general difficult to get an explicit
expression for this canonical equivalence. In this letter we shall adopt an
alternative approach to improve this situation. We shall consider the canonical
transformation of the CGHS model introduced in \cite{cangemi} and reinterpret
it as a B\"acklund transformation. In this new context we shall be able to
generalize this B\"acklund transformation for a generic model of dilaton
gravity. The B\"acklund transformation will define then an explicit canonical
transformation.\\

The canonical transformation for the CGHS theory proposed in \cite{cangemi}
makes use of the following auxiliary canonical variables $\e^0$, $\e^1$,
$p_0$, $p_1$ defined by
\bea
\label{ben1}
r^a & = & \frac{1}{\sqrt{2}} \eta^a \, , \\
\label{ben2}
\pi_a & = & \sqrt{2} (p_a -\frac{1}{2} \epsilon_{ab} \eta^{b\prime}) \, ,
\eea
where $\epsilon_{ab}$ is the antisymmetric tensor with $\epsilon_{01}=-1$. In
terms of the canonical variables $(\eta^a,p_a)$ the constraints have the form
\bea
\label{hf}
H & = & \e^{0\prime} p_1 + \e^{1\prime} p_0 - (p_1^2-p_0^2) \, , \\
\label{pf}
P & = & \e^{0\prime} p_0 + \e^{1\prime} p_1 \, .
\eea
The canonical transformation is then defined by the following relations
\bea
\label{cjz1}
\e^0 & = & \frac{1}{2\l} e^{-\r} \left( \pr \sinh \t - 2\f^{\prime} \cosh \t
\right) \, , \\
\label{cjz2}
\e^1 & = & \frac{1}{2\l} e^{-\r} \left( \pr \cosh \t - 2\f^{\prime} \sinh \t
\right) \, , \\
\label{cjz3}
p_0 & = & \phantom{-} 2\l e^{\r} \sinh \t \, , \\
\label{cjz4}
p_1 & = & - 2\l e^{\r} \cosh \t \, ,
\eea
where $\t=\frac{1}{2} \int_{-\infty}^{x} d\tilde{x} \pf$. This transformation
is canonical because it can be obtained from a generating functional. It is
interesting to point out now that the above field redefinition can
be regarded as a B\"acklund transformation connecting the dilaton-gravity
equations (\ref{eq1}), (\ref{eq2}) for the CGHS model with free field equations
\be
\pmas \pmen \e^0 = 0 = \pmas \pmen \e^1 \, .
\ee

We want now to generalize the above transformation for a generic model. It is
easy to see that a transformation of the form
\bea
\e^{0\prime} & = & \tilde{F}^{-1} \left[ (H + \tilde{F}^2) \cosh \t
+ P \sinh \t \right] \, , \\
\e^{1\prime} & = & \tilde{F}^{-1} \left[ (H + \tilde{F}^2) \sinh \t
+ P \cosh \t \right] \, , \\
p_0 & = & -\tilde{F} \sinh \t \, , \\
p_1 & = & \phantom{-} \tilde{F} \cosh \t \, ,
\eea
where $\tilde{F}$ and $\t$ are arbitrary functions, bring the constraints of a
generic theory to the form (\ref{hf}), (\ref{pf}). Because the canonicity of
the transformation requires that the fields $\e^0$ and $\e^1$ verify free field
equations, a natural generalization of the B\"acklund transformation defined by
(\ref{cjz1})-(\ref{cjz4}) is given by the following ansatz
\bea
\label{bac1}
\e^{0\prime} & = & -\frac{1}{2\l} F^{-1} e^{-\r} \left[ (H + 4\l^2 F^2
e^{2\r}) \cosh \t + P \sinh \t \right] \, , \\
\label{bac2}
\dot{\e}^1 & = & -  2\l F e^{\r} \cosh \t \, ,
\eea
where
\bea
\label{def1}
H(\r,\f) & = & -2 \dot{\f} \dot{\r} + 2(\f^{\prime\prime} -
\f^{\prime} \r^{\prime}) - 4\l^2 V(\f) e^{2\r} \, , \\
\label{def2}
P(\r,\f) & = & -2( \dot{\f} \r^{\prime} - \dot{\f}^{\prime} + \f^{\prime}
\dot{\r}) \, , \\
\label{def3}
F(\r,\f) & = & \exp \left\{ \frac{\l^2}{2} \partial^{-1}_+ \partial^{-1}_- 
\left( \frac{d}{d\f}V(\f) e^{2\r} \right) \right\} \, , \\
\label{def4}
\t(\r,\f) & = & -\int_{-\infty}^{x} d\tilde{x} (\dot{\r} + \frac{\dot{F}}{F})
\, .
\eea
The main goal of this letter is the following result:\\

\noindent {\it The transformation defined by (\ref{bac1}), (\ref{bac2}) is a
B\"acklund transformation that relates the solutions $\r$, $\f$ of a generic
dilaton gravity theory and the solutions $\e^0$, $\e^1$ of a free field
theory}.\\

\noindent {\it Proof:} First of all, we observe the relation that there
exists between the functions $F$ and $\t$. Taking derivatives in expression
(\ref{def3}) we can rewrite (\ref{eq1}) as
\be
\label{free}
\pmas \pmen (\r + \ln F) = 0 \, .
\ee
Using (\ref{def4}) we then obtain the following identities
\bea
\label{id1}
\pmas (\r + \ln F) & = & - \pmas \t \, , \\
\label{id2}
\pmen (\r + \ln F) & = & \phantom{+} \pmen \t \, .
\eea
Now we take derivatives in (\ref{bac1}), (\ref{bac2})
\bea
& \pmas \pmen \e^{0\prime} = \frac{-1}{2\l} F^{-1} e^{-\r} \left\{ 
\pmas \pmen P \sinh \t + \pmas \pmen H \cosh \t \right. + & \nonumber \\
& \left[ F^2 e^{2\r} \left( \pmas (\r + \ln F) \pmen \t +
\pmen (\r + \ln F) \pmas \t + \pmas \pmen \t \right) \right. + & \nonumber \\
& \pmas H \pmen \t + \pmen H \pmas \t - \pmas P
\pmen (\r + \ln F) - \pmen P \pmas (\r + \ln F) - & \nonumber \\
& P \left( \pmas \pmen (\r + \ln F) - \pmas (\r + \ln F)
\pmen (\r + \ln F) - \pmas \t \pmen \t \right) - & \nonumber \\
& \left. H \left( \pmas (\r + \ln F) \pmen \t + \pmen (\r + \ln F)
\pmas \t - \pmas \pmen \t \right) \right] \sinh \t + & \nonumber \\
& \left[ F^2 e^{2\r} \left( \pmas \pmen (\r + \ln F) + \pmas (\r + \ln F)
\pmen (\r + \ln F) + \pmas \t \pmen \t \right) \right. + & \nonumber \\
& \pmas P \pmen \t + \pmen P \pmas \t - \pmas H
\pmen (\r + \ln F) - \pmen H \pmas (\r + \ln F) - & \nonumber \\
& H \left( \pmas \pmen (\r + \ln F) - \pmas (\r + \ln F)
\pmen (\r + \ln F) - \pmas \t \pmen \t \right) - & \nonumber \\
& \left. \left. P \left( \pmas (\r + \ln F) \pmen \t + \pmen (\r + \ln F)
\pmas \t - \pmas \pmen \t \right) \right] \cosh \t \right\} \, , & \\
& \pmas \pmen \dot{\e}^1 = -2\l F e^{\r} \left\{ \left[ \pmas (\r + \ln F)
\pmen \t + \pmen (\r + \ln F) \pmas \t + \pmas \pmen \t \right] \cosh \t +
\right. & \nonumber \\
& \left. \left[ \pmas \pmen (\r + \ln F) + \pmas (\r + \ln F) \pmen
(\r + \ln F) + \pmas \t \pmen \t \right] \sinh \t \right\} \, , &
\eea
and taking into account (\ref{free}), (\ref{id1}), (\ref{id2}) the above
expressions become
\bea
\label{arr}
\pmas \pmen \e^{0\prime} & = & \frac{-1}{2\l} F^{-1} e^{-\r} \left\{
\pmas \pmen P \sinh \t + \pmas \pmen H \cosh \t + \right. \nonumber \\
& & \left[ \pmas (H - P) \pmen \t + \pmen (H + P) \pmas \t \right] \sinh \t +
\nonumber \\
& & \left. \left[ \pmas (P - H) \pmen \t + \pmen (P + H) \pmas \t
\right] \cosh \t \right\} \, , \\
\pmas \pmen \dot{\e}^1 & = & 0 \, .
\eea
Finally, using the Bianchi identities $\partial_{\pm} (H \mp P)=0$, we see
that the r.h.s. of (\ref{arr}) vanishes and then
\bea
\label{f1}
\pmas \pmen \e^0 & = & 0 \, , \\
\label{f2}
\pmas \pmen \e^1 & = & 0 \, .
\eea
Moreover, the above derivation also work on the other way around, so if $\e^0$,
$\e^1$ satisfy the free field equations (\ref{f1}), (\ref{f2}) then $\r$,
$\f$ satisfy the equations of motion (\ref{eq1}), (\ref{eq2}).\\

To construct the fully generalized canonical transformation we introduce the
cano\-ni\-cally conjugated momenta $\pr=-2\dot{\f}$, $\pf=-2\dot{\r}$,
$p_0=\dot{\e}^0$, $p_1=\dot{\e}^1$ as independent variables. Then we get
\bea
\label{gt1}
\e^{0\prime} & = & \frac{-1}{2\l} F^{-1} e^{-\r} \left[ (H + 4\l^2 F^2
e^{2\r}) \cosh \t + P \sinh \t \right] \, , \\
\label{gt2}
\e^{1\prime} & = & \frac{-1}{2\l} F^{-1} e^{-\r} \left[ (H + 4\l^2 F^2
e^{2\r}) \sinh \t + P \cosh \t \right] \, , \\
\label{gt3}
p_0 & = & \phantom{-} 2\l F e^{\r} \sinh \t \, , \\
\label{gt4}
p_1 & = & -  2\l F e^{\r} \cosh \t \, ,
\eea
where $H$, $P$ are given by
\bea
\label{h}
H & = & -\frac{1}{2}\pr\pf + 2(\f^{\prime\prime} -\r^{\prime}\f^{\prime}) -
4\l^2 V(\f) e^{2\r} \, , \\
\label{p}
P & = & \pr \r^{\prime} - \pr^{\prime} + \pf \f^{\prime} \, ,
\eea
$F$ is given by (\ref{def3}) and $\t = \frac{1}{2} \int_{-\infty}^{x}
d\tilde{x} (\pf - 2\frac{\dot{F}}{F})$. We have also seen that this
transformation maps the constraints (\ref{h}), (\ref{p}) into the previous free
form (\ref{hf}), (\ref{pf}). It is well known that a B\"acklund transformation
can be viewed as a canonical transformation. This is so because there are no
other expressions for the Poisson brackets that reproduce the Hamiltonian
equations of motion for the free fields $\eta^a$, $p_a$.\\

The CJZ transformation for the CGHS model is recovered when $F(\r,\f)=1$.
Then $\t=\frac{1}{2} \int_{-\infty}^{x} d\tilde{x} \pf$ and (\ref{gt1})-
(\ref{gt4}) read as
\bea
\label{c1}
\e^{0\prime} & = & \frac{1}{2\l} e^{-\r} \left[ (\frac{1}{2}\pr\pf -
2(\f^{\prime\prime} +\r^{\prime}\f^{\prime})) \cosh \t -
(\pr \r^{\prime} - \pr^{\prime} + \pf \f^{\prime}) \sinh \t \right] \\
\label{c2}
\e^{1\prime} & = & \frac{1}{2\l} e^{-\r} \left[ (\frac{1}{2}\pr\pf -
2(\f^{\prime\prime} +\r^{\prime}\f^{\prime})) \sinh \t -
(\pr \r^{\prime} - \pr^{\prime} + \pf \f^{\prime}) \cosh \t \right] \\
p_0 & = & \phantom{-} 2\l e^{\r} \sinh \t \, , \\
p_1 & = & -  2\l e^{\r} \cosh \t \, .
\eea
It is easy to see that (\ref{c1}),(\ref{c2}) leads to (\ref{cjz1}),
(\ref{cjz2}).\\

For the model with an exponential (Liouville) potential $V=e^{\b \f}$ the
function $F$ is also local. The transformations (\ref{gt1})-(\ref{gt4}) with
$F=e^{-\b \f}$ provides an alternative canonical transformation for the induced
2D Polyakov gravity, which differs from the one obtained by using the classical
solutions \cite{cruz3}. Another interesting example is the Jackiw-Teitelboim
model ($V(\f)=\f$). In this case one cannot find a local expression for the 
function $F$. However, taking into account that $\r$ verifies a Liouville
equation the field $\r + \ln F$ coincide with the well known free field 
associated, via a canonical transformation, to a Liouville field
\cite{hoker}.\\

In this letter we have constructed a B\"acklund transformation relating a
generic 2D dilaton-gravity model to a generally covariant free field theory.
This way we have provided an explicit canonical transformation connecting both
theories.

\section*{Acknowledgements}

D. J. Navarro acknowledges the Ministerio de Educaci\'on y Cultura for a FPI
fellowship. We thank J. Cruz for dicussions.


\end{document}